\documentclass[conference]{IEEEtran}
\usepackage[utf8]{inputenc}
\usepackage{graphicx}
\usepackage{hhline}
\usepackage{amsmath}
\usepackage{amssymb}
\usepackage{amsfonts}
\usepackage{mathtools}
\usepackage{epsfig}
\usepackage{epstopdf}
\usepackage[justification=centering]{caption}
\usepackage{subcaption}
\usepackage{stfloats}
\usepackage{cite}
\usepackage[CJKbookmarks=true,
bookmarksnumbered=true,
bookmarksopen=true]{hyperref}
\hypersetup{hidelinks}

\def\degree{${}^{\circ}$}
\title{Modeling and Measurements for Multi-path Mitigation with Reconfigurable Intelligent Surfaces}
\vspace{-4em}
\author{\IEEEauthorblockN{Ruya Zhou\IEEEauthorrefmark{1}, Xiangyu Chen\IEEEauthorrefmark{1},  Wankai Tang\IEEEauthorrefmark{1}, Xiao Li\IEEEauthorrefmark{1}, Shi Jin\IEEEauthorrefmark{1}, Ertugrul~Basar\IEEEauthorrefmark{2}, Qiang Cheng\IEEEauthorrefmark{3}, Tie Jun Cui\IEEEauthorrefmark{3}}
	\IEEEauthorblockA{\IEEEauthorrefmark{1}National Mobile Communications Research Laboratory, Southeast University, Nanjing 210096, P. R. China}
	\IEEEauthorblockA{\IEEEauthorrefmark{2}Department of Electrical and Electronics Engineering, Koc University, Istanbul 34450, Turkey}
	\IEEEauthorblockA{\IEEEauthorrefmark{3}State Key Laboratory of Millimeter Waves, Southeast University, Nanjing 210096, China}
\vspace{-3em}

}	
\date{}
\begin{document}
	\captionsetup{font={small}}

\maketitle

\begin{abstract}
A reconfigurable intelligent surface (RIS) is capable of manipulating electromagnetic waves with its flexibly configurable unit cells, thus is
an appealing technology to resist fast fading caused by multi-path in wireless communications.
In this paper, a two-path propagation model for RIS-assisted wireless communications is proposed by considering both the direct path from the transmitter to the receiver and the assisted path provided by the RIS.
The proposed propagation model unveils that the phase shifts of RISs can be optimized by appropriate configuration for multi-path fading mitigation.
In particular, four types of RISs with different configuration capabilities are introduced and their performances on improving received signal power in virtue of the assisted path to resist fast fading are compared through extensive simulation results.
In addition, an RIS operating at 35 GHz is used for experimental measurement. The experimental results verify that an RIS has the ability to combat fast fading and thus improves the receiving performance, which may lay a foundation for further researches.
\end{abstract}

\vspace{-0.5cm}
\section{Introduction}

\begin{figure*}[t]
	\centering
	\vspace{-0.8cm}
	\begin{center}	
		\centering		
		\includegraphics[height=4.8cm,width=9.8cm]{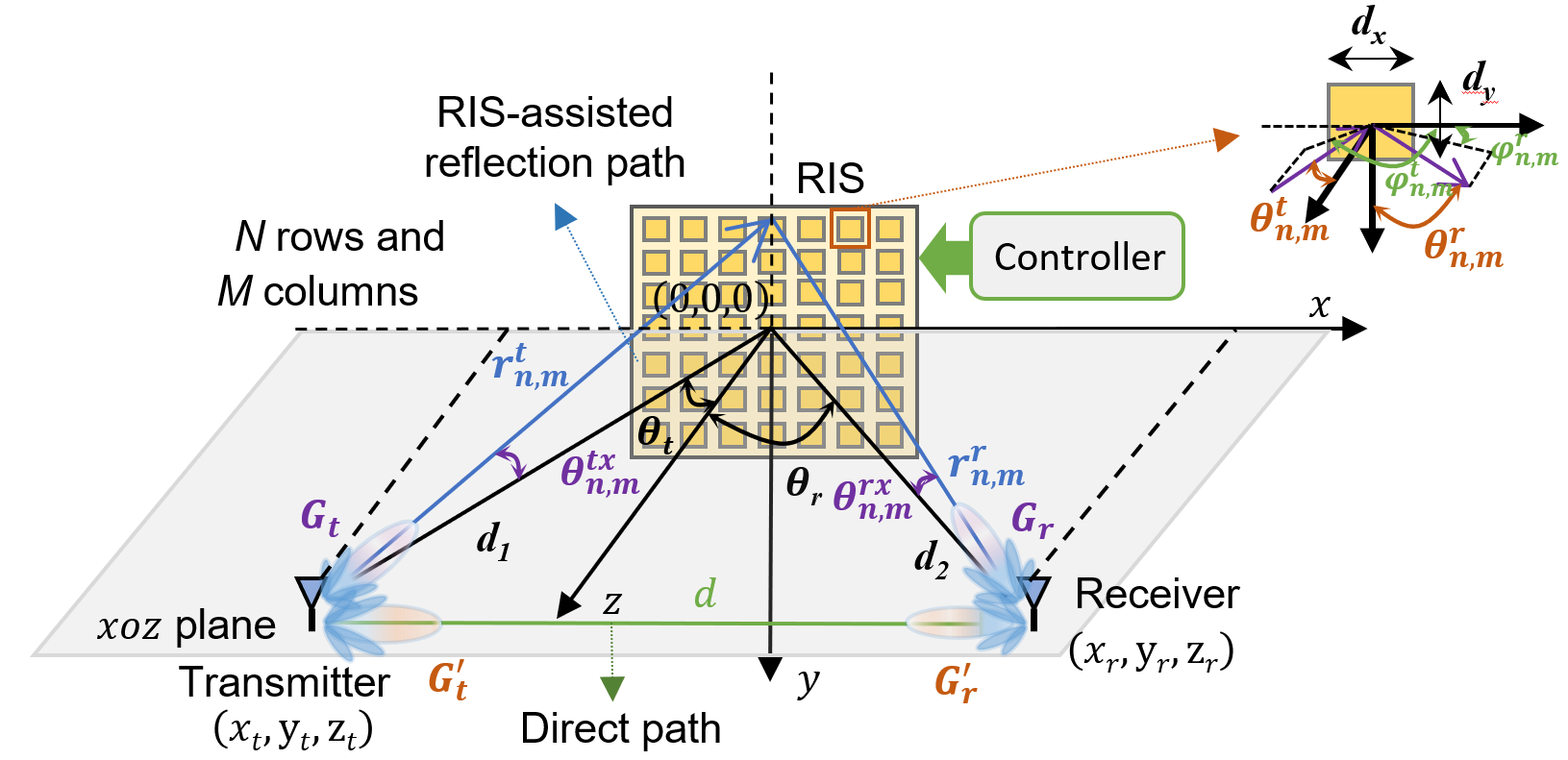}			
		\caption{Diagram of the two-path propagation model for an RIS-assisted wireless communication system.}
		\label{fig:1}
		\vspace{-0.7cm}
	\end{center}
\end{figure*}

As a new enabling technology of 6G,
the reconfigurable intelligent surface (RIS)-empowered communication has drawn increasingly high attention
due to its superior capability to manipulate electromagnetic waves in the field of wireless communications \cite{cui2014coding}.
Studies on RISs of various aspects are being actively carried out in recent years.
By modeling the RIS as a diagonal phase-shift matrix in the signal model,
the transmission performance is optimized in RIS-assisted wireless communication systems in \cite{huang2019reconfigurable,wu2019intelligent,han2019large}.
In \cite{dai2020realization} and \cite{tang2020mimo}, RISs are utilized to realize direct information modulation in real time. Besides, RISs are applied for beamforming in RIS-based wireless communication prototypes in \cite{dai2020reconfigurable} and \cite{arun2020rfocus}.
Moreover, previous works \cite{tang2021wireless,ellingson2019path,di2020analytical,danufane2021path,tang2021path,wang2021received,basar2021reconfigurable} theoretically model the wireless channels of RIS-assisted communication systems. In particular, the path loss characteristics of RIS-assisted wireless communications are experimentally measured in \cite{tang2021wireless,tang2021path} and \cite{wang2021received} by using different RISs in both sub-6 GHz and millimeter wave bands.
However, studies in previous works on modeling and measuring of multi-path fading are relatively rare and the performance degradation caused by fast fading through multi-path propagation at the receiver is not unveiled through experimental measurements so far.
The phase shifts of RISs can be designed to enhance the received signal power due to their flexibly configurable unit cells, thus making RISs potentially applied to mitigate the effects brought about by multi-path fading \cite{basar2019reconfigurable,zhang2021spatial}.

Motivated by the above conditions, a two-path propagation model for RIS-assisted wireless communications considering both the direct path and the RIS-assisted path is proposed
to reveal the fast fading caused by multi-path propagation in free-space. Then the improvement on receiving performance after introducing RISs is discussed.
We analyze the received signal power by numerical simulations with the assistance of four types of RISs, which are distinct with different configurable degrees of freedom. Moreover, fast fading is revealed through experimental results. We further configure the phase shifts of RISs to resist fast fading and the measurement results are in good agreement with the modeling results.

The remainder of this paper is arranged as follows. Section II proposes the two-path propagation model for RIS-assisted wireless communications.
Section III formulates the received signal power and
shows the numerical simulation results of RISs with different configurable capabilities on multi-path mitigation.
Section IV reports the experimental results that validate the ability of RISs to combat fast fading. And conclusion is drawn in Section V.

\vspace{-0.1cm}
\section{System Model}
\vspace{-0.1cm}

In this section, the description of the two-path propagation model for RIS-assisted wireless communications is presented.
As shown in Fig. \ref{fig:1}, a general two-path propagation model for an RIS-aided single-input single-output (SISO) wireless communications is considered. An RIS with $N$ rows and $M$ columns is deployed between the transmitter and the receiver to form a two-path propagation scenario,
which includes the direct path and
the assisted path composed of $M{\times}N$ sub-paths.
Taking the geometric center of the RIS as the origin, the spatial coordinate system is established and the RIS is located on the $xoy$ plane.
The elevation angle (relative to the z-axis) and azimuth angle from the center of the RIS to the transmitter are respectively $\theta_t$ and $\phi_t$.
Also, the elevation and azimuth angles from the center of the RIS to the receiver are respectively $\theta_r$ and ${\varphi}_r$.
The transmitter and the receiver are placed on the $xoz$ plane, thus ${\varphi}_t=180^{\circ}$ and ${\varphi}_r=0$\degree \cite{tang2021wireless
}.

In the two-path propagation model for RIS-assisted wireless communications, $d$ denotes the distance between the transmitter and the receiver. $d_1$ and $d_2$ denote the distances from the center of the RIS to the transmitter and the receiver, respectively.
$U_{n,m}$ represents the unit cell which is in the $n^{th}$ row and $m^{th}$ column.
The physical dimensions of each unit cell are $d_x$ and $d_y$ in $x$ and $y$ directions, respectively.
The reflection coefficient of each unit cell is ${\Gamma}_{n,m}=A_{n,m} e^{j\varphi_{n,m}} $, where $A_{n,m}$ denotes the reflection amplitude and $\varphi_{n,m}$ denotes the reflection phase. The central coordinate position of $U_{n,m}$ is $((\frac{M+1}{2}-m) d_x$, $(\frac{N+1}{2}-n) d_y, 0)$, where $m\in[1,M]$, and $n\in[1,N]$. $M$ and $N$ are assumed to be even numbers.
The elevation and azimuth angles from the $U_{n,m}$  to the transmitter are $\theta_{n,m}^t$ and $\varphi_{n,m}^t$, respectively.  Also, the elevation and azimuth angles from the $U_{n,m}$ to the receiver are $\theta_{n,m}^r$ and $\varphi_{n,m}^r$, respectively.
$F(\theta_{n,m},\varphi_{n,m})$ is the normalized power radiation pattern of each unit cell, which reveals the relation between the incident / reflected power density and the incident / reflected angle.
The distances from the $U_{n,m}$ to the transmitter and the receiver are $r_{n,m}^t$ and $r_{n,m}^r$, respectively.

Transmitter transmits signals with normalized power radiation pattern $F^{tx} (\theta_{n,m}^{tx},\varphi_{n,m}^{tx})$ as well as antenna gains $G_t$ and $G_t'$ to two directions, and the transmitting power is $P_t$.
Signals are received by the receiver whose receiving antenna has normalized power radiation pattern $F^{rx} (\theta_{n,m}^{rx},\varphi_{n,m}^{rx})$ and antenna gains $G_r$ and $G_r'$. 
Among them, the elevation angle and azimuth angle from the transmitting antenna to the $U_{n,m}$ are $\theta_{n,m}^{tx}$ and $\varphi_{n,m}^{tx}$, respectively. The elevation angle and azimuth angle of the receiving antenna to the $U_{n,m}$ are respectively $\theta_{n,m}^{rx}$ and $\varphi_{n,m}^{rx}$.

\vspace{-0.1cm}
\section{Theoretical Modeling of Multi-path Propagation and Mitigation with RIS}
In this section, the explicit expression of the received signal power for the proposed model in Section II is given and four types of RISs with different configurable degrees of freedom are introduced. Moreover, a comparison is made among different types of RISs on the performance of resistance to fast fading.

\vspace{-0.1cm}
\subsection{Explicit Expression of the Received Signal Power}

\begin{table}
	\vspace{0.36cm}
	\centering
	\caption{Comparison of four different RISs.}
	{\begin{tabular}[l]{|c|c|c|c|}			
			\hline			
			Type & Configurable degrees of freedom & Cost & CSI \\
			\hline			
			RIS1  & 0\degree or 180\degree/integral configuration & low  & not necessary \\
			\hline
			RIS2 & full 360\degree/integral configuration & moderate & not necessary \\
			\hline
			RIS3  & 0\degree or 180\degree/cell configuration &  moderate   & not necessary\\
			\hline
			RIS4 & full 360\degree/cell configuration & high &  necessary\\
			\hline			
	\end{tabular}}
	\vspace{-0.67cm}
	\label{symbols:1}	
\end{table}

The following theorem gives the relation between the received signal power and the key parameters in the two-path propagation model for RIS-assisted wireless communications.

\textbf{Theorem 1}. The received signal power of the receiver in the two-path propagation model for RIS-assisted wireless communications can be expressed as

\vspace{-0.4cm}
\begin{equation}
	\begin{aligned}
		&P_r=\bigg|\sum_{m=1}^{M}\sum_{n=1}^{N}d_xd_y\Gamma_{n,m}\frac{\sqrt{P_tG_tG_rF_{n,m}^{combine}}}{4{\pi}r_{n,m}^tr_{n,m}^r}&&\\
		&e^{-j\frac{2\pi(r_{n,m}^t+r_{n,m}^r)}{\lambda}}+
		\frac{\lambda\sqrt{P_tG_t'G_r'}}{4\pi{d}}e^{-j\frac{2\pi{d}}{\lambda}}\bigg|^2,&&\\
	\end{aligned}
\end{equation}
where $F_{n,m}^{combine}$ can be written as \cite{tang2021path}

\vspace{-0.36cm}
\begin{equation}
	\begin{aligned}
		&F_{n,m}^{combine}=F_{n,m}^{tx}(\theta_{n,m}^{tx},\varphi_{n,m}^{tx})
		F_{n,m}^{t}(\theta_{n,m}^{t},\varphi_{n,m}^{t})&&\\
		&F_{n,m}^{r}(\theta_{n,m}^{r},\varphi_{n,m}^{r})
		F_{n,m}^{rx}(\theta_{n,m}^{rx},\varphi_{n,m}^{rx}).&&\\
	\end{aligned}
\end{equation}

Proof: See Appendix A.

Theorem 1 reveals that the two-path propagation model consists of a direct path and an RIS-assisted path that includes $M\times{N}$ sub-paths. Signal superposition of multi-path components leads to phase congruency or phase cancellation, which may result in fast fading at the receiver.

\begin{figure}[b]
	\vspace{-0.6cm}
	\setlength{\abovecaptionskip}{-0.02cm}
	\setlength{\belowcaptionskip}{-0.4cm}
	\centering
	\includegraphics[height=4.6cm,width=8.2cm]{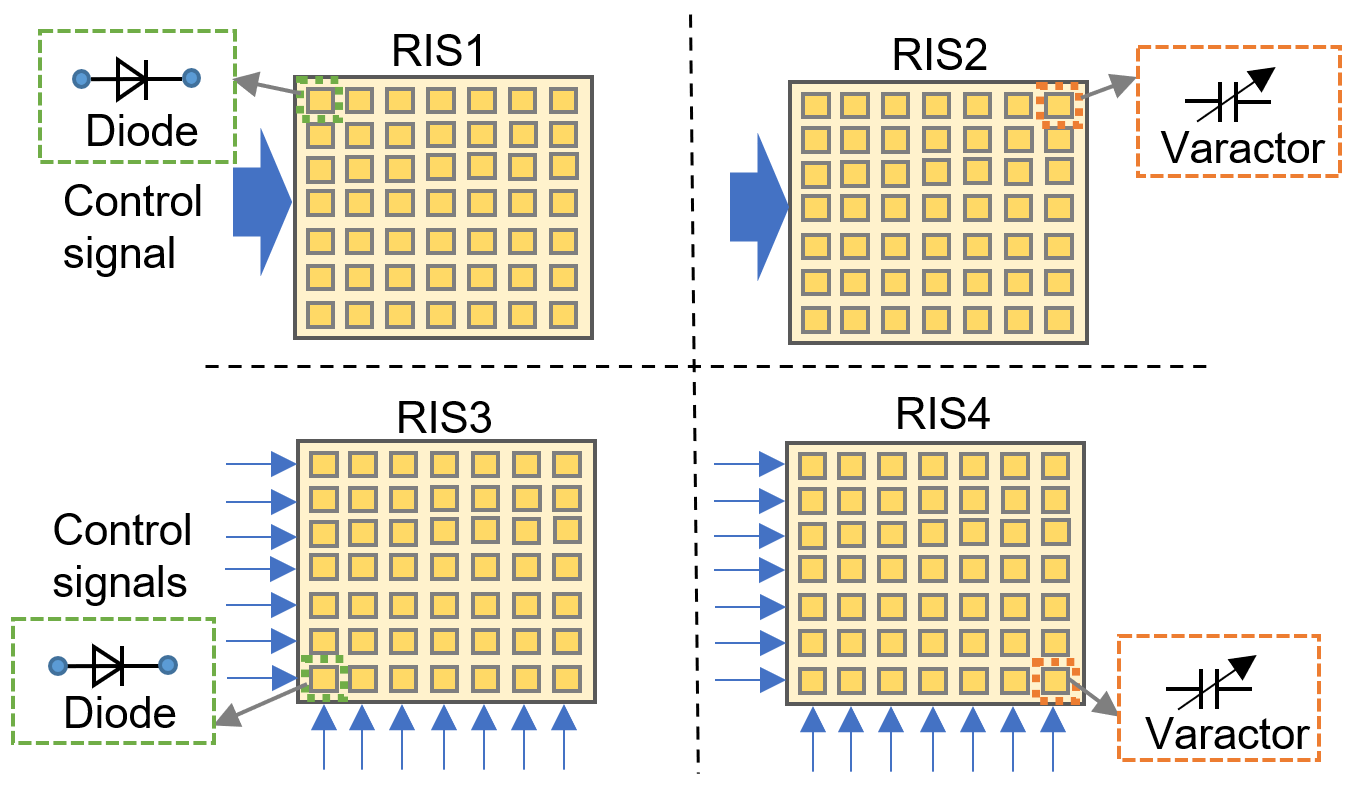}
	\caption{Four types of RISs with different configurable capabilities.}
	\label{fig:2}
\end{figure}

\vspace{-0.2cm}
\subsection{Comparison of Received Signal Power with Different RISs}

According to different configurable degrees of freedom, RISs can be divided into four types: (a) integrally and discretely configurable unit cells; (b) integrally and continuously configurable unit cells; (c) individually and discretely configurable unit cells; (d) individually and continuously configurable unit cells. The integral configuration uses one control signal to control the whole RIS, while the cell configuration uses multiple signals to control each unit cell independently.
The unit cells that use diodes can realize discrete configuration with 0\degree and 180\degree phase shifts, while the unit cells that use varactors can realize continuous configuration with a range of full 360\degree phase shifts.
For ease of exposition, the four types of RISs above are referred to as RIS1, RIS2, RIS3, and RIS4, respectively.
The details are shown in Fig. \ref{fig:2} and Table \ref{symbols:1}.

\begin{figure*}[t]
	\centering
	\vspace{-0.9cm}
	\begin{center}
		\begin{minipage}[c]{0.5\textwidth}		
			\centering		
			\includegraphics[height=6.5cm,width=8.2cm]{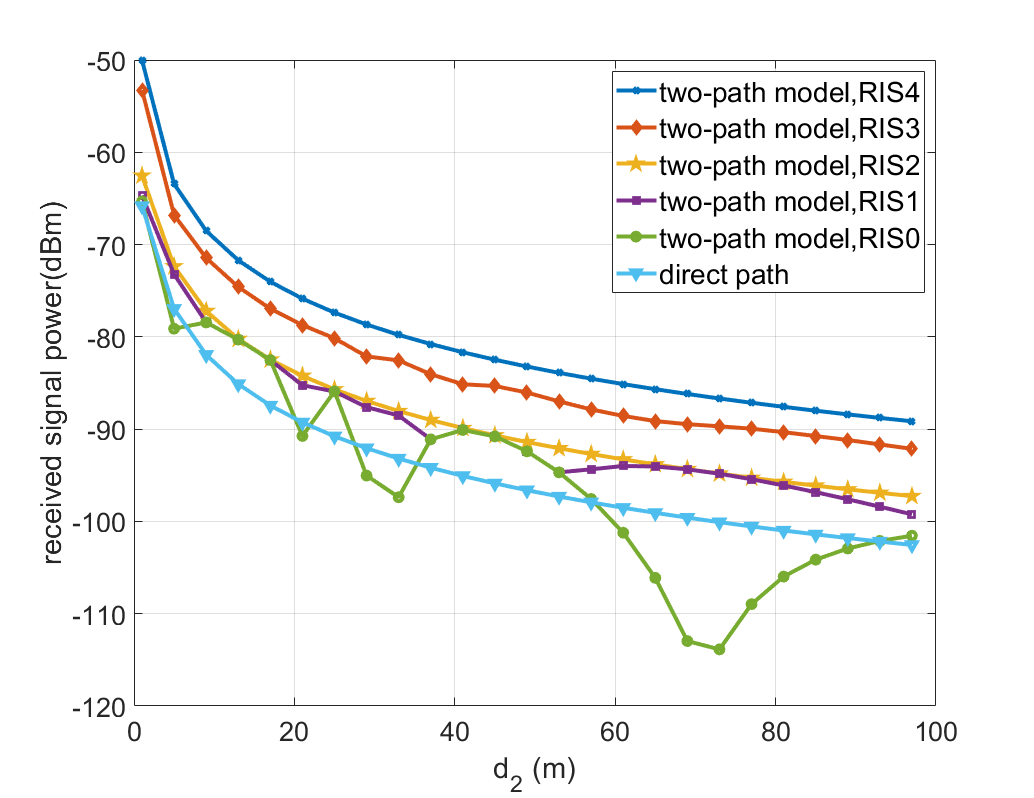}	
			\subcaption{$\theta_t=\theta_r=45$\degree}		
		\end{minipage}%
		\begin{minipage}[c]{0.5\textwidth}		
			\centering		
			\includegraphics[height=6.5cm,width=8.2cm]{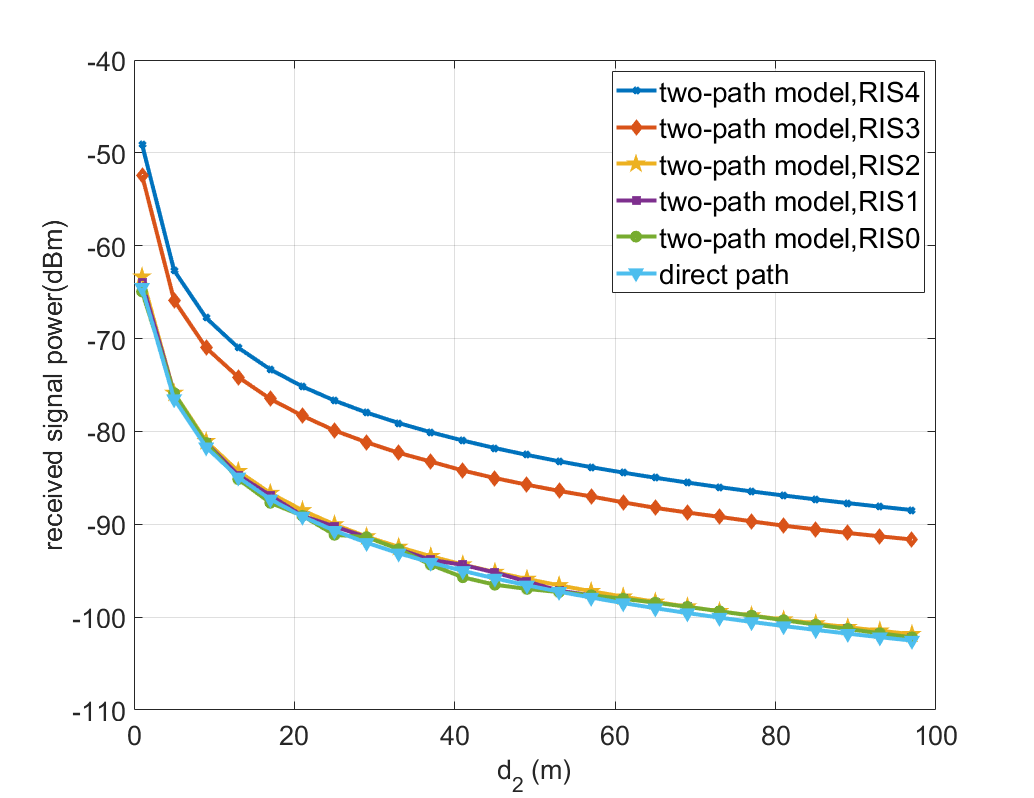}	
			\subcaption{$\theta_t=30$\degree, $\theta_r=45$\degree}		
		\end{minipage}
		\caption{Comparison of different RISs for fast fading suppression to enhance received signal power.}
		\label{fig:3}
		\vspace{-0.83cm}
	\end{center}
\end{figure*}

RISs with different configurable capabilities differ in ability to combat fast fading, which can be compared through received signal power.
The simulation parameters are given as follows.
The numbers of rows and columns of the unit cells are equal, i.e., $M = N=64$.
The distance $d_1$ from the transmitter to the center of the RIS is fixed to $1$ m and the distance $d_2$ from the RIS to the receiver changes from $1$ m to $100$ m.
$P_t = 0 $ dBm, $G_t=G_t'=G_r=G_r'=1$, $|\Gamma_{n,m}|=|\Gamma|=0.8$, $d_x=d_y=3.8 $ mm, $F_{n,m}^{tx}=F_{n,m}^{rx}=1$, $F_{n,m}^{t}=\cos\theta_{n,m}^t$, $F_{n,m}^{r}=\cos\theta_{n,m}^r$, $\theta_t=\theta_r=45$\degree.
RIS0 is configured as an isophase surface, i.e., the reflection coefficient $\Gamma_ {n, m}$ of each unit cell is the same. RIS0 without phase optimization is introduced here to compare with the four types of RISs with phase optimization.

As can be seen from Fig. \ref{fig:3}(a), the receiving performance with RIS4 whose unit cells can be individually and continuously configured is better than others.
This is as expected since the configurable degree of freedom of RIS4 is the highest.
The channel state information (CSI) is necessary to design phase shifts on RIS4 for each path and each unit cell can be configured to the corresponding optimal phase shifts to reach the maximum value of the received signal power.
Therefore, RIS4 can be utilized to effectively resist fast fading thus increasing the received signal power compared with RIS0, whose curve exhibits many fluctuations.

The second best option is RIS3 being individually and discretely configured.
Although RIS3 cannot be configured at a range of 360\degree,
voting algorithm \cite{arun2020rfocus} can be adopted for cell configuration
to get better performance without the need for CSI. By randomly selecting the state of each unit cell: 0 \degree or 180 \degree, several different configurations of the phase distribution of RIS3 are generated by iterating for many times. The optimal configuration of phase shifts is selected, and then the received signal power under the corresponding configuration is calculated, which is the maximum power that can be achieved when unit cells of RIS are being discretely configured.
Even if the unit cells of RIS3 cannot be continuously configured, the received signal power at the receiver is significantly higher than RIS0.

The third highest value of the received signal power is provided by RIS2. The maximum value can be obtained by traversing 360 \degree of all phases at each distance.
The limit of RIS2 is the phases of all unit cells must be the same, so the configurable flexibility is relatively low, and the increase of received signal power is limited. However, the receiving performance can be improved to a certain degree.
Therefore, even if the phase of each unit cell cannot be configured individually, we can utilize the RIS for integral configuration to improve the received signal power.

The worst one in the four cases is RIS1, in which the configurable degree of freedom is the lowest, thus the received signal power is only improved a little than RIS0 and the direct path.
Compared with other curves, the curve of the received signal power of RIS1 is not smooth since the configurable flexibility of integral configuration is relatively low and only two states can be selected in the optimization.
When the RIS can only be configured as a whole, the receiving performance can be slightly improved by designing discrete phases.

In conclusion, the four types of RISs are all capable of combating fast fading by reducing fluctuations and improving received signal power compared with RIS0. Moreover, the higher the configurable degree of freedom, the higher the received signal power can be received, and vice versa.

\begin{figure*}[t]
	\centering
	\vspace{-0.5cm}
	\begin{center}
		\begin{minipage}[c]{0.5\textwidth}		
			\centering		
			\includegraphics[height=5.4cm,width=8.2cm]{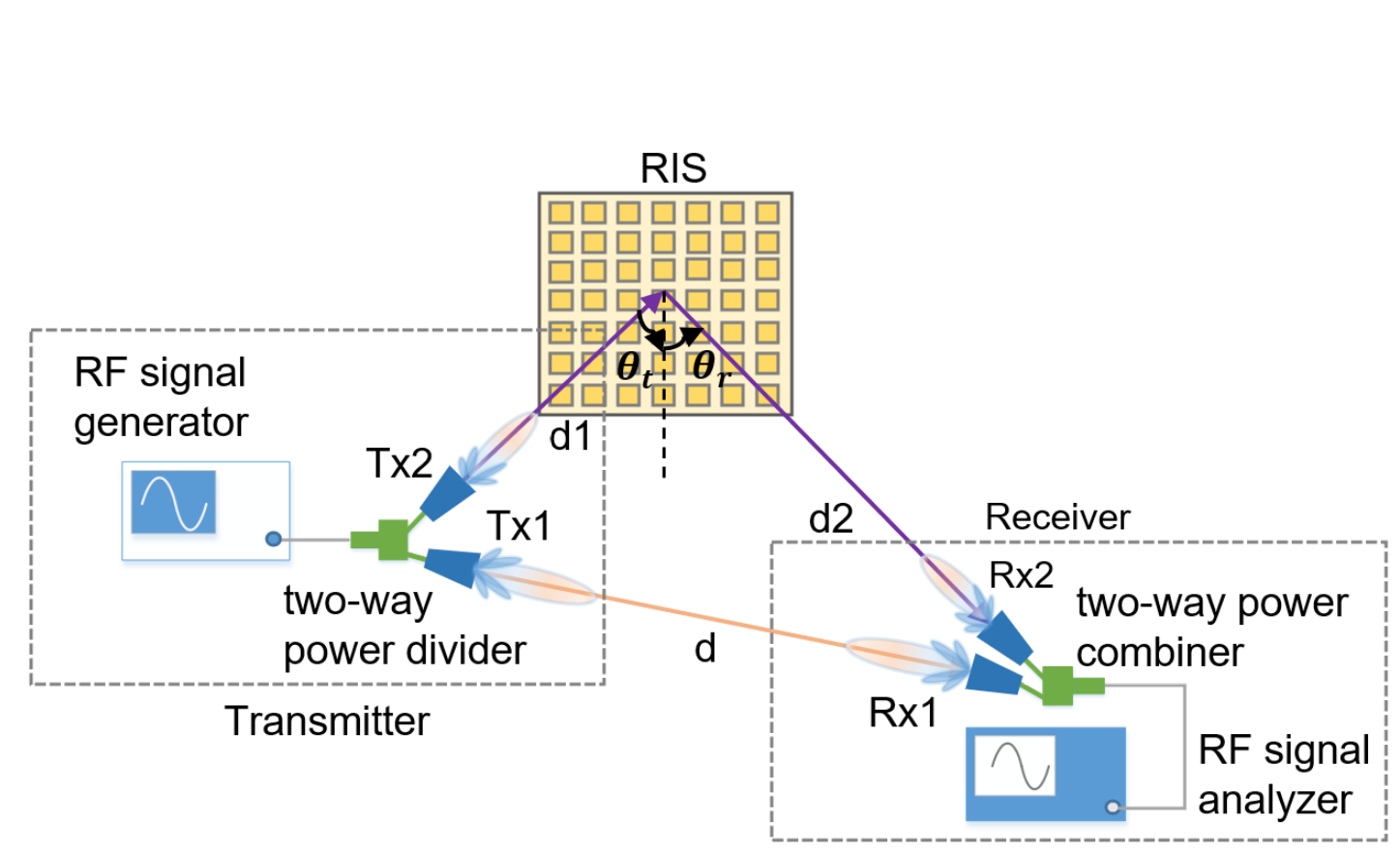}	
			\subcaption{Diagram}		
		\end{minipage}%
		\begin{minipage}[c]{0.5\textwidth}		
			\centering		
			\includegraphics[height=5.4cm,width=8.2cm]{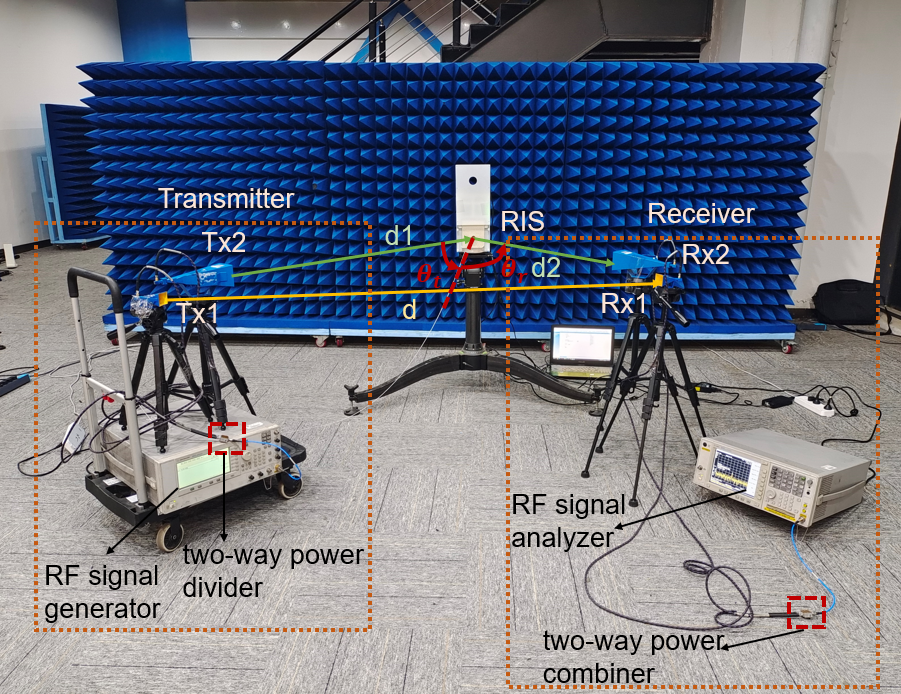}	
			\subcaption{Photo}		
		\end{minipage}
		\caption{Experimental measurement setup.}
		\label{fig:4}
		\vspace{-1.1cm}
	\end{center}
\end{figure*}

It is noted that the restrictive condition for the above conclusion is that the angle of arrival is equal to that of departure ($\theta_t=\theta_r=45$\degree). The reason for the good performance of integral control for RIS1 and RIS2 is that they are equal-phase surfaces and can only perform specular reflection.
In the following, we alter the angle of arrival to 30\degree, making it different from the angle of departure in order to make a more general conclusion.
As depicted in Fig. \ref{fig:3}(b), RIS4 and RIS3 still keep good performance. However, integral configuration methods for RIS1 and RIS2 no longer enhance the received signal power compared with RIS0. In this case, cell configuration reflects sharp superiority over integral configuration.

To sum up, through the above simulations, the receiving performance of RIS3 and RIS4 is quite satisfying, since enough degree of freedom in configuration is provided. On the other hand, the discrete configuration (0\degree or 180\degree) is relatively simple compared with the full 360\degree configuration, thus making discrete and cell configuration (RIS3) a good choice in practical applications.

\vspace*{-0.15cm}
\section{Measurement Validation of the Multi-path Mitigation with RIS}

The experimental measurements are carried out to further validate the proposed two-path propagation model for RIS-assisted wireless communications and prove that RISs have the ability to combat fast fading.

\vspace{-0.1cm}
\subsection{Experiment setup}

The fabricated RIS utilized in the measurement is phase-programmable with 1-bit coding, which has the same configurable degrees of freedom with RIS1. The detailed parameters of this RIS are reported in Table \ref{symbols:2}.

\begin{table}	
	\centering
	\vspace{0.34cm}
	\caption{Parameters of the RIS and antennas.}
	\vspace{-0.02cm}	
	{\begin{tabular}[l]{|c|c|}			
			\hline
			$M/N$ & 30\\
			\hline
			$d_x/d_y$ & 3.8 mm\\
			\hline
			$|\Gamma_{n,m}|$ & 0.8 \\
			\hline
			$\angle\Gamma_{n,m}$ & 0\degree, when coding ``0"; 180\degree, when coding ``1"\\
			\hline			
			$f$  & 35 GHz \\
			\hline			
			$G_t$(RIS-assisted path) & 323.6 (25.1dB)\\
			\hline
			$G_r$(RIS-assisted path) & 323.6 (25.1dB)\\
			\hline
			$G_t'$(direct path) & 128.8 (21.1dB)\\
			\hline
			$G_r'$(direct path) & 128.8 (21.1dB)\\
			\hline
			$F^{tx}(\theta,\phi)$ & $(cos\theta)^{161}, \theta\in[0,\pi/2]; 0, \theta\in(\pi/2,\pi]$\\
			\hline
			$F^{rx}(\theta,\phi)$ & $(cos\theta)^{161}, \theta\in[0,\pi/2]; 0, \theta\in(\pi/2,\pi]$\\
			\hline			
	\end{tabular}}
    \vspace{-0.6cm}
	\label{symbols:2}	
\end{table}

The transmitter mentioned in Section II contains an RF signal generator (Agilent E8257D), a two-way power divider, and two Tx horn antennas called Tx1 and Tx2, respectively. And the receiver consists of a spectrum analyzer (Agilent E4447A), a two-way power combiner, and two Rx horn antennas named Rx1 and Rx2, respectively. The parameters of the used antennas are also listed in Table \ref{symbols:2}.

Fig. \ref{fig:4} illustrates our measurement setup, which is composed of an RIS, a transmitter, a receiver, and other accessories such as cables.
The RIS is located vertically and the transmitter is placed ahead of the RIS with a distance of $d_1=1$ m. The arrival angle is $\theta_t=45$\degree. We constantly move the receiver by changing $d_2$ from $0.6$ m to $3$ m at the direction of $\theta_r=45$\degree. Other system parameters, such as $\theta_{n,m}^{t}$ and $\varphi_{n,m}^{t}$ are also determined by the positions of the transmitter, the receiver and the RIS. We record the data at every $0.2$ m.
The RF signal generator provides the RF signal with a constant power of $15$ dBm to Tx1 and Tx2 through the two-way power divider.
One transmits through Tx1 to the Rx1 directly and the other travels through Tx2 which illuminates the RIS and then is reflected to Rx2. The RF signal analyzer gathers signals received by antennas Rx1 and Rx2 through the power combiner.

\begin{figure}[t]
	\centering
	\vspace{0.2cm}
	\includegraphics[scale=0.3]{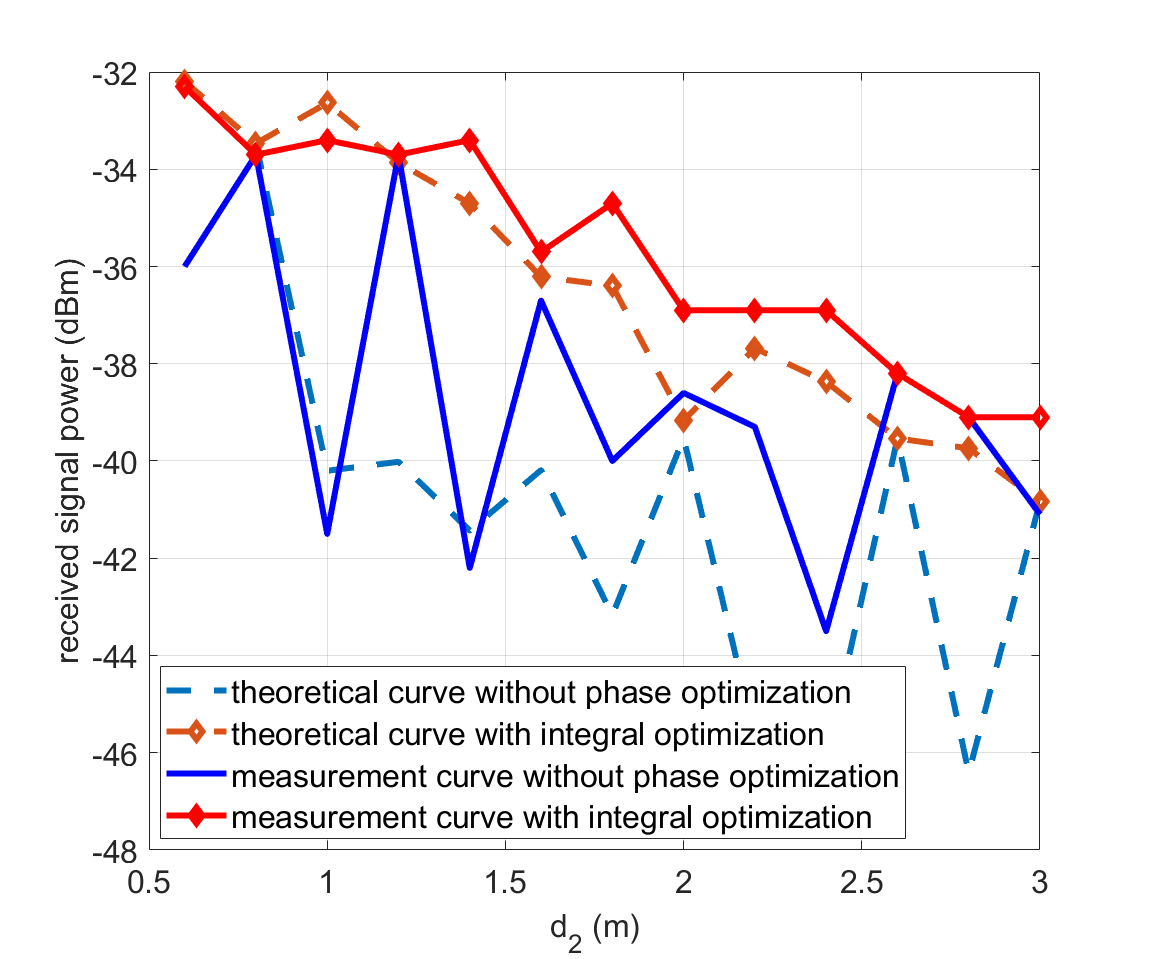}
	\caption{Measurement results of multi-path mitigation with the RIS.}
	\vspace{-0.6cm}
	\label{fig:5}
\end{figure}

\vspace{-0.2cm}
\subsection{Measurement Results}
At first, no phase optimization is applied to RIS to better observe the multi-path fading phenomenon at the receiver. Therefore, all the unit cells of the RIS in our experimental measurements are all set to an identical reflection coefficient of $|\Gamma_{n,m}| = 0.8$, $\angle\Gamma_{n,m} = 0$\degree
at the beginning. In this case, RIS serves as an isophasic surface to reflect the incident signal in a specular way without any phase optimization, aiming at revealing fast fading caused by multi-path transmission. It is worth noting that there are cable losses in our experiments, so we make a calibration in our simulation to compensate the power loss introduced by RF cables \cite{tang2021path}.
Fig. \ref{fig:5} sketches the comparison of received signal power versus $d_2$ between experimental one and theoretical one, from which we can see that the downward and fluctuating trend of the measured curve of the RIS without phase optimization is in good agreement with the theoretical one which clearly reveals fast fading.

Then an experiment utilizing RIS is carried out to prove that RISs have the ability to mitigate negative impact brought by fast fading and enhance the received signal power at the receiver. There are only two phase shifts ($\angle\Gamma_{n,m}=$0\degree/$\angle\Gamma_{n,m}=$180\degree) in the integral configuration for the RIS to maximize the received signal power at different distances of $d_2$. As depicted in Fig. \ref{fig:5}, although the configurable degree of freedom of RIS is relatively low, the fast fading at the receiver is mitigated to some extent and an obvious improvement of the received signal power can be seen from the measurement with integral configuration compared with that without phase configuration, which is close to the theoretical results. It is fully proved that RISs can be used to combat fast fading to enhance the received signal power by configuring the phase shifts of each unit cell.

\vspace{-0.08cm}
\section{Conclusion}
\vspace{-0.08cm}

In this paper, a two-path propagation model for RIS-assisted wireless communications has been proposed. The general formula of received signal power has been formulated. We have proved that RISs can be utilized to mitigate fast fading caused by multi-path propagation due to the flexibly configured phase shifts on their surfaces.
Depending on the configurable degrees of freedom, RISs can be configured to achieve different receiving performances. A comparison of the received signal power has been made among different types of RISs with different configurable capabilities.
Moreover, practical measurements of the two-path propagation model for RIS-assisted in wireless communications have been conducted to reveal fast fading. Also, an RIS has been utilized in measurements to prove its ability in multi-path mitigation. This work may guide researchers
in analyzing and simulating the performance of various RIS-assisted wireless communication systems under multi-path effects.

\begin{appendices}
\section{Proof of Theorem 1}


In the two-path propagation model for RIS-aided wireless communications, the received signal intensity reflected by each unit cell $U_ {n,m}$ to the receiver can be expressed as \cite{tang2021path}

\vspace{-0.2cm}
\begin{equation}
S_{n,m}^r=d_xd_y\Gamma_{n,m}\frac{\sqrt{P_tG_tG_rF_{n,m}^{combine}}}{4\pi{r_{n,m}^t}{r_{n,m}^r}}e^{-j\frac{2\pi(r_{n,m}^t+r_{n,m}^r)}{\lambda}}.
\end{equation}


The total received signal intensity that reaches the receiver through the RIS-assisted path can be written as
\vspace{-0.11cm}
\begin{equation}
S^r=\sum_{m=1}^M{\sum_{n=1}^N{S_{n,m}^r}}.
\end{equation}




The received signal intensity of the direct path is

\begin{equation}
S_{LOS}=\sqrt{P_{LOS}}e^{-j2\pi{d}},
\end{equation}
where $P_{LOS}$ is the received signal power of the direct path.

$P_{LOS}$ can be expressed as
\begin{equation}
	P_{LOS}=\frac{P_tG_t'G_r'{\lambda}^2}{({4\pi{d}})^2}.
\end{equation}

The total received signal power can be expressed as

\begin{equation}
P_{r}=|S^r+S_{LOS}|^2.
\end{equation}

Theorem 1 can be obtained by substituting (3)-(6) into (7).

\end{appendices}

\bibliographystyle{ieeetr}
\bibliography{refer1}

\end{document}